\documentclass[lettersize,journal]{IEEEtran}
\usepackage{amsmath,amsfonts}
\usepackage{algorithmic}
\usepackage{algorithm}
\usepackage{array}
\usepackage[caption=false,font=normalsize,labelfont=sf,textfont=sf]{subfig}
\usepackage{textcomp}
\usepackage{stfloats}
\usepackage{url}
\usepackage{verbatim}
\usepackage{graphicx}
\usepackage{xcolor}
\usepackage{booktabs}           
\usepackage{multirow}         
\usepackage[table,xcdraw]{xcolor}
\usepackage{cite}
\usepackage[hidelinks]{hyperref}
\usepackage{adjustbox}
\usepackage{colortbl}
\usepackage{subcaption}
\usepackage{float}
\usepackage{placeins}
\usepackage{tcolorbox}
\usepackage{tablefootnote}
\usepackage{tabularx}
\definecolor{mypink}{RGB}{255,230,230}
\definecolor{lightred}{RGB}{255, 230, 230}
\definecolor{lightblue}{RGB}{230, 240, 255}

\usepackage{cite}
\begin{document}
\raggedbottom
\title{Language Models as Semantic Teachers: Post-Training Alignment for Medical Audio Understanding}

\author{Tsai\hbox{-}Ning~Wang,
        Lin\hbox{-}Lin~Chen,
        Neil~Zeghidour,
        and Aaqib~Saeed%
\thanks{Tsai‐Ning Wang and Lin‐Lin Chen are with
Eindhoven University of Technology, The Netherlands (e-mail: t.n.wang@tue.nl; l.chen@tue.nl).}%
\thanks{Neil Zeghidour is with Kyutai, France (e-mail: neil@kyutai.org).}%
\thanks{Aaqib Saeed is with the 
Eindhoven University of Technology, The Netherlands and the Eindhoven Artificial Intelligence Systems Institute, The Netherlands.
(e-mail: a.saeed@tue.nl).}%
}

% \markboth{Journal of \LaTeX\ Class Files,~Vol.~xx, No.~x, Dec~2025}%

\maketitle

\begin{abstract}
Pre-trained audio models excel at detecting acoustic patterns in auscultation sounds but often fail to grasp their clinical significance, limiting their use and performance in diagnostic tasks. To bridge this gap, we introduce AcuLa (Audio–Clinical Understanding via Language Alignment), a lightweight post-training framework that instills semantic understanding into any audio encoder by aligning it with a medical language model, which acts as a ``semantic teacher.'' To enable alignment at scale, we construct a large-scale dataset by leveraging off-the-shelf large language models to translate the rich, structured metadata accompanying existing audio recordings into coherent clinical reports. Our alignment strategy combines a representation-level contrastive objective with a self-supervised modeling, ensuring that the model learns clinical semantics while preserving fine-grained temporal cues. AcuLa achieves state-of-the-art results across 18 diverse cardio-respiratory tasks from 10 different datasets, improving the mean AUROC on classification benchmarks from 0.68 to 0.79 and, on the most challenging COVID-19 cough detection task, boosting the AUROC from 0.55 to 0.89. Our work demonstrates that this audio-language alignment transforms purely acoustic models into clinically-aware diagnostic tools, establishing a novel paradigm for establishing a paradigm for injecting clinical-language semantics into audio representations for audio-based health monitoring.

\end{abstract}


\begin{figure*}[t]
    \centering    
    \includegraphics[width=0.9\textwidth]{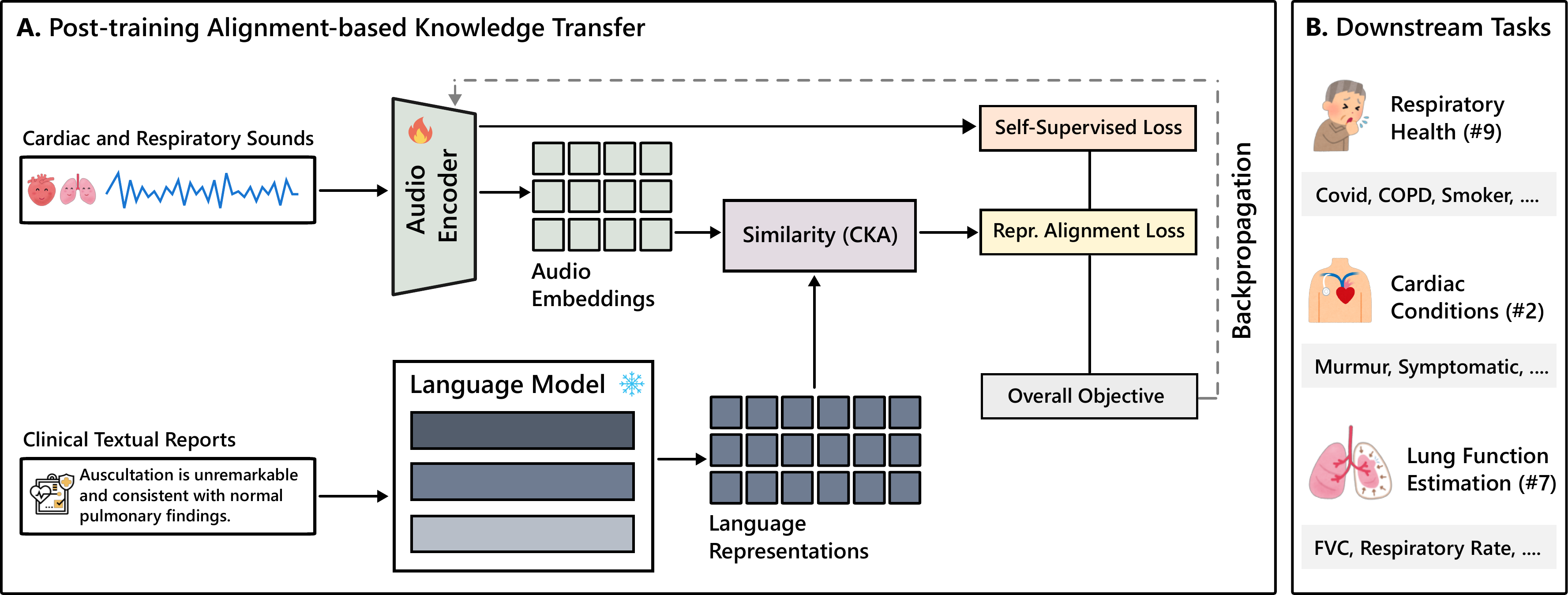}
    \caption{\small{Architecture of the audio-language alignment framework. (A) Audio encoders extract features from clinical recordings, which are aligned with language representations via similarity matching. (B) Down‑stream tasks enabled by the aligned model, including (i) respiratory‑health classification (9 tasks), (ii) cardiac‑condition detection (2 tasks) and (iii) lung‑function estimation (7 tasks).}}
    \label{fig:arch}
\end{figure*}

\section{Introduction}
Existing audio encoders capture subtle temporal and spectral patterns in auscultation sounds but still lack explicit clinical semantics. This leaves them ``semantically blind,'' limiting their use in high-stakes diagnostic tasks. A fundamental paradox follows: while large language models (LLMs) understand medical concepts such as ``systolic murmurs'' and ``wheezes,'' this knowledge remains disconnected from the audio models that process raw signals. Digital stethoscopes and other sensors can capture rich acoustic data, but without a bridge to semantic meaning, much of this information remains underused.

\looseness=-1
Multimodal contrastive learning, popularized by frameworks such as CLIP \cite{radford2021learning}, seeks to bridge such divides by aligning heterogeneous modalities in a shared embedding space. This has enabled strong cross-modal retrieval and classification, but these methods often suffer from a persistent ``modality gap,'' where embeddings from different modalities form distinct clusters, limiting fine-grained alignment and interpretability \cite{liang2022mind, ray2023cola}. This issue is especially critical in clinical settings, where subtle acoustic differences can carry major diagnostic significance and require precise semantic grounding.

Existing approaches have addressed this gap through architectural changes \cite{chen2023revisiting}, auxiliary objectives \cite{lee2022uniclip}, or post-training alignment \cite{yamaguchi2025post}, but they mainly focus on aligning two perceptual modalities. Even recent knowledge transfer methods such as \cite{gan2025cross} follow this paradigm, improving language models with knowledge from vision models. These approaches assume that knowledge flows from concrete perception to abstract representation. In contrast, our work frames the problem as directed, asymmetric knowledge infusion. We leverage the broad semantic knowledge of an LLM as a ``semantic teacher'' to guide and enrich a specialized ``acoustic student.'' This introduces a distinct challenge: grounding high-level clinical concepts from text into the fine-grained temporal patterns of raw audio, which remains largely unexplored.

This frontier is especially important for domains rich with temporal and semantic information, such as medical audio. Millisecond-scale events like the onset of a lung crackle or the specific timing of a heart murmur contain precise clinical information that current audio-only models struggle to link to a diagnosis. To address this, we introduce AcuLa (Audio–Clinical Understanding via Language Alignment), a general, post-training framework that instills clinical semantic understanding into any pre-trained audio encoder. In our approach, a frozen, LLM acts as a ``semantic teacher,''  guiding an audio ``student'' model to map acoustic patterns to their corresponding clinical meanings. 

We demonstrate AcuLa's effectiveness in the challenging domain of cardio-respiratory health. By synthetically generating a large-scale dataset of clinical reports from structured metadata, we create the necessary paired data to align audio recordings with their clinical interpretations. Our results show that this alignment transforms standard audio encoders into clinically-aware models that can better differentiate subtle pathologies and significantly improves downstream tasks performance.

Our work makes the following key contributions:

\begin{itemize}
\item \textbf{Model-Agnostic Audio-Language Knowledge Transfer:} We propose a general framework (AcuLa) to enhance pre-trained audio encoders by transferring knowledge from LLMs. This demonstrates a novel paradigm where LLMs serve as semantic teachers for specialized auditory models (see Figure~\ref{fig:arch} for an overview).
\item \textbf{Preservation-Focused Teacher-Student Design:} Our lightweight architecture connects pre-trained models through minimal projection layers, preserving specialized knowledge in both models while enabling efficient cross-modal knowledge transfer without expensive retraining or architectural modifications.

\item \textbf{Synthetic Data Generation from Structured Metadata:} We construct paired audio-text data by generating clinical auscultation reports ($\approx$100,000) from real metadata of audio clip using powerful off-the-shelf LLMs, producing semantically accurate and diverse narratives aligned with each recording.
\item \textbf{Dual Objective Optimization:} We combine a representation alignment loss with audio self-supervised modeling (such as masked acoustic reconstruction loss) in a multi-task setting. This dual objective ensures that the model learns semantic relationships while maintaining the fine-grained temporal precision essential for medical audio analysis.
\end{itemize}

\section{Related Work}
\label{sec:related}

\subsection{Medical Audio Analysis}
Respiratory and cardiac sound analysis has traditionally been treated as a unimodal task. Supervised deep learning performs well with sufficient expert labels, while self-supervised pre-training helps in low-resource settings \cite{zhang2024towards}. More recent work has begun incorporating text; for example, RespLLM \cite{zhang2024respllm} combines spectral features with clinical notes through cross-modal attention. However, these methods usually process audio and text separately and fuse them only at a late stage for prediction. This limits the audio encoder’s ability to learn semantically rich representations. In contrast, our work aligns audio and language at the feature level to directly ground acoustic events in clinical meaning.

\subsection{Cross-Modal Alignment}
Bridging the semantic gap between modalities is a central challenge in machine learning. Existing methods can be broadly understood through the lens of their alignment strategies.

\textbf{Representation Alignment.} The dominant paradigm, popularized by CLIP \cite{radford2021learning}, learns a shared embedding space where corresponding pairs from different modalities are projected to be close. This contrastive approach has been successfully extended to the audio domain with models like CLAP \cite{elizalde2023clap} and AudioCLIP \cite{guzhov2022audioclip}, enabling better cross-modal learning. However, these frameworks often struggle with a ``modality gap,'' where representations remain clustered by their original modality, hindering fine-grained understanding \cite{liang2022mind}. This limitation is particularly critical for medical signals, where subtle pattern differences are diagnostically vital.

\textbf{Knowledge Transfer and Distillation.} Alignment can also be viewed as directed knowledge transfer. Generative methods such as AudioLM \cite{borsos2023audiolm} and AudioGen \cite{kreuk2022audiogen} learn audio conditioned on text, implicitly inducing shared structure. Knowledge distillation \cite{hinton2015distilling} transfers representations from a ``teacher'' to a ``student,'' typically in supervised, task-specific settings. Most related to our work is regularization-based alignment, where one representation space is regularized to match another (e.g., CMAR \cite{gan2025cross} regularizes an LLM using features from a vision model). Connector-based speech--text approaches \cite{tan-etal-2025-ssr, yu-etal-2023-speech} instead model token-level correspondences by training cross-modal adapters end to end and often updating the language model. In contrast, we perform post-training \emph{global} embedding alignment: the medical LLM is frozen, and lightweight projection heads align audio encoder representations using schema-constrained, clip-level textual supervision.

\textbf{Our Contributions and Positioning.} Despite these advances, two gaps remain. First, semantic alignment between medical audio representations and clinical text is underexplored. Second, cross-modal knowledge transfer has mostly focused on static modalities (e.g., vision--language), with limited attempts to unify audio (e.g., ImageBind~\cite{girdhar2023imagebind}). AcuLa bridges these gaps by introducing a lightweight framework for representation alignment tailored to temporal medical audio, using a pre-trained LLM as a semantic teacher for adapting a specialized audio encoder. The frozen LLM provides clinical-language semantics as a regularizer; we do not claim physiological ground truth, causal understanding, or clinical decision-making.

\section{Methodology}
\label{sec:method}
Our framework, \textbf{AcuLa}, establishes a post-training alignment between a pre-trained audio encoder and a pre-trained language model. We achieve this by introducing lightweight, trainable projection heads and fine-tuning the audio encoder, guided by a dual objective that promotes semantic similarity while retaining fine-grained acoustic detail. The core language model remains frozen, acting as a fixed semantic teacher.

\subsection{Problem Statement}
Let $\mathcal{A}_\theta$ be a pre-trained audio encoder with parameters $\theta$, and $\mathcal{L}_\phi$ be a pre-trained language model with parameters $\phi$. Given a batch of $B$ paired examples $\{(\mathbf{x}_i, \mathbf{r}_i)\}_{i=1}^B$, where $\mathbf{x}_i \in \mathbb{R}^{T \times F}$ is an audio spectrogram and $\mathbf{r}_i$ is the corresponding textual report, our goal is to learn parameters for an audio projection head $P^{\text{audio}}_{\psi_a}$ and a language projection head $P^{\text{language}}_{\psi_l}$. Let $\psi = \{\psi_a, \psi_l\}$ be the set of all trainable projection parameters. These heads map the outputs of their respective encoders into a shared $d$-dimensional embedding space:
\begin{align}
    \mathbf{h}^{\text{audio}}_i &= P^{\text{audio}}_{\psi_a}(\mathcal{A}_\theta(\mathbf{x}_i)) \in \mathbb{R}^d \\
    \mathbf{h}^{\text{language}}_i &= P^{\text{language}}_{\psi_l}(\mathcal{L}_\phi(\mathbf{r}_i)) \in \mathbb{R}^d
\end{align}
The learning objective is to optimize the audio encoder parameters $\theta$ and the projection parameters $\psi$ such that (i) the representations $\mathbf{h}^{\text{audio}}_i$ and $\mathbf{h}^{\text{language}}_i$ for corresponding pairs are semantically aligned, while (ii) the audio encoder's ability to model detailed acoustic patterns is preserved. The LLM parameters $\phi$ remain frozen throughout.

\begin{figure}[ht]             
    \centering
    \includegraphics[width=\columnwidth]{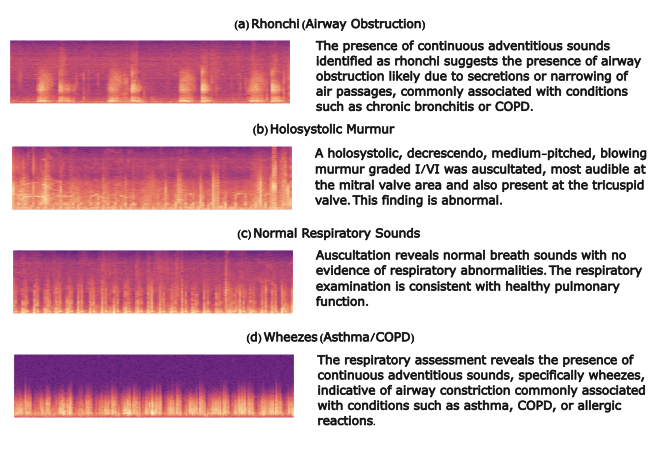}
    \captionsetup{font=small}  
    \caption{Spectrograms of cardiopulmonary sounds with paired clinical reports. (a) Rhonchi showing continuous adventitious sounds from airway obstruction. (b) Holosystolic murmur indicating mitral valve pathology. (c) Normal breath sounds with clear pulmonary function. (d) Wheezes revealing airway constriction associated with asthma or COPD.}
    \label{fig:spec}
\end{figure}

\subsection{Alignment Architecture}

Our architecture (Figure \ref{fig:arch}.A) is designed to be lightweight and preservation-focused. The core components are the audio encoder $\mathcal{A}_\theta$ and the language model $\mathcal{L}_\phi$. The knowledge transfer is mediated by two simple, trainable projection heads, $P^{\text{audio}}_{\psi_a}$ and $P^{\text{language}}_{\psi_l}$, implemented as multi-layer perceptrons (MLPs). This design enables efficient alignment while minimizing disruption to the pre-trained models architectures.

\subsection{Training Objective}
Our multi-task training objective is designed to simultaneously achieve semantic alignment and preserve the temporal fidelity of the audio encoder. The full objective is a weighted sum of two losses:
\begin{equation}
\mathcal{L}(\theta, \psi) = \lambda_{\text{align}} \mathcal{L}_{\text{align}}(\theta, \psi; \phi) + \lambda_{\text{SSM}} \mathcal{L}_{\text{SSM}}(\theta)
\label{eq:overall_loss}
\end{equation}
Here, we optimize the audio encoder parameters $\theta$ and the projection head parameters $\psi$, while the LLM parameters $\phi$ remain frozen. Unless stated otherwise, we set $\lambda_{\text{align}}=\lambda_{\text{SSM}}=1.0$.

\textbf{Semantic Alignment via Centered Kernel Alignment (CKA).} To align the two modalities, we require a robust similarity metric between the sets of batch embeddings, $\mathbf{H}^{\text{audio}} = [\mathbf{h}^{\text{audio}}_1, \dots, \mathbf{h}^{\text{audio}}_B]^T$ and $\mathbf{H}^{\text{language}} = [\mathbf{h}^{\text{language}}_1, \dots, \mathbf{h}^{\text{language}}_B]^T$. We obtain $\mathbf{h}^{\text{language}}$ by mean-pooling the last-layer hidden states of the frozen LLM, and map both modalities to a shared embedding space via two MLP projectors.
We use Centered Kernel Alignment (CKA) \cite{kornblith2019similarity}, a metric that compares the geometric structure of representation spaces, making it invariant to isotropic scaling and rotation. CKA is defined via the Gram matrices of the mean-centered representations, $\mathbf{G}^{\text{audio}} = \bar{\mathbf{H}}^{\text{audio}T} \bar{\mathbf{H}}^{\text{audio}}$ and $\mathbf{G}^{\text{language}} = \bar{\mathbf{H}}^{\text{language}T} \bar{\mathbf{H}}^{\text{language}}$:
\begin{equation}
\mathcal{A}(\mathbf{H}^{\text{audio}}, \mathbf{H}^{\text{language}}) = \frac{\langle \mathbf{G}^{\text{audio}}, \mathbf{G}^{\text{language}} \rangle_F}{\|\mathbf{G}^{\text{audio}}\|_F \|\mathbf{G}^{\text{language}}\|_F}
\end{equation}
where $\langle \cdot, \cdot \rangle_F$ is the Frobenius inner product. The alignment loss seeks to maximize this similarity:
\begin{equation}
\mathcal{L}_{\text{align}}(\theta, \psi; \phi) = 1 - \mathcal{A}(\mathbf{H}^{\text{audio}}, \mathbf{H}^{\text{language}})
\end{equation}
We use the LLM as a source of \emph{clinical-language semantics} for regularization and do not interpret this alignment as physiological ground truth or causal biomedical understanding. The alignment is performed on clip-level global embeddings and does not explicitly model event-level temporal--semantic correspondence within a recording.

\textbf{Acoustic Preservation via Self-Supervised Modeling (SSM).} The alignment loss alone may cause the audio encoder to discard acoustic information not captured by simplified text reports (representation collapse \cite{jing2021understanding}). We therefore include the audio encoder's self-supervised objective, $\mathcal{L}_{\text{SSM}}$. In our implementation, $L_{SSM}$ is instantiated as the audio encoder's masked acoustic reconstruction loss. We use the encoder's default masked reconstruction setup: the mel-spectrogram is split into $4\times4$ patches, $70\%$ are masked uniformly at random. For many state-of-the-art audio models, this is a masked acoustic modeling loss~\cite{huang2022masked} (or a contrastive objective~\cite{saeed2021contrastive}), acting as a regularizer to preserve acoustic modeling capability. The final optimization thus becomes:
\begin{equation}
\theta^*, \psi^* = \arg\min_{\theta, \psi} \left[ \lambda_{\text{align}} \mathcal{L}_{\text{align}}(\theta, \psi; \phi) + \lambda_{\text{SSM}} \mathcal{L}_{\text{SSM}}(\theta) \right]
\end{equation}
This dual objective balances the acquisition of new semantic knowledge with the preservation of existing acoustic capabilities.

\subsection{Synthetic Alignment Data Generation} 
A significant challenge in medical multimodal learning is the scarcity of large-scale, paired audio-text datasets. To overcome this, we devise a scalable strategy to generate high-quality clinical text from existing audio datasets with structured metadata. We leverage an off-the-shelf LLM, GPT-4o \cite{openai2024gpt4technicalreport}, to synthesize clinical reports. For each audio recording from public datasets like ICBHI \cite{DVN/HT6PKI_2023} and Circor \cite{oliveira2021circor}, we compile its available metadata—patient demographics, recording conditions, and diagnostic labels (e.g., presence of crackles/wheezes)—into a structured prompt (see Appendix \ref{apd:prompt} and Figure \ref{fig:spec}). The LLM is tasked to act as a clinical specialist (e.g., a pulmonologist) and generate a concise, natural-language report based \textit{only} on the provided information. With precise and explicit prompting of LLM, we ensure factual grounding while encouraging linguistic diversity. Examples of metadata–report pairs are provided in Appendix~\ref{apd:meta-report-examples}. This process yields a corpus of over $100,000$ paired audio-report samples, summarized in Table \ref{tab:dataset_stats}.

\begin{table}[H]
\caption{Statistics of the our synthesized data used for model alignment. For \textbf{UK Covid-19}, \textbf{IC+EX} denotes the union of \emph{Induced Cough} (IC) and \emph{Exhalation} (EX). Duration values show average duration in seconds.}
\centering
% \footnotesize
% \setlength{\tabcolsep}{4pt}
\resizebox{\columnwidth}{!}{%
\begin{tabular}{l|l|c|c|c}
\toprule
\textbf{Dataset} & \textbf{Modality} & \textbf{\#Reports} & \textbf{\#Audio} & \textbf{Avg. Duration (s)} \\
\midrule
ICBHI      & Lung sound & 6,899  &  6,899 & 22.2 \\
HFLung     & Lung sound & 9,765  & 10,554 & 15.0 \\
UK Covid-19 & IC+EX      & 72,999 & 40,252 &  5.9 \\
CoughVID   & Induced cough & 11,314 &  7,179 &  6.9 \\
Circor     & Heart sound & 1,568 &  5,282 & 22.87 \\
SPRSound   & Lung sound  & 1,772 &  2,496 & 11.15 \\
ZCHSound   & Heart sound & 1,259 &  1,259 & 20.06 \\
\bottomrule
\end{tabular}
}
\label{tab:dataset_stats}
\end{table}

\subsection{Expert Verification of Synthetic Reports}
\label{sec:expert_verification}

To validate the LLM-generated synthetic reports used for alignment, we conducted an expert review of 50 metadata--report pairs sampled from the synthesized corpus (10 from each of 5 datasets), covering diverse respiratory and cardiac scenarios. Each pair included the structured metadata and its generated clinical report. A pulmonologist evaluated metadata consistency, unsupported added detail, and clinical plausibility.
We define \emph{accuracy} as the proportion of reports fully consistent with the source metadata, and unsupported-detail rate as the proportion containing information not directly supported by the metadata. Results are summarized in Table~\ref{tab:expert_verification}.

As shown in Table~\ref{tab:expert_verification}, all 50 reviewed reports were fully consistent with the source metadata and clinically plausible. Unsupported added detail appeared in 6 cases (12\%), but these did not contradict the metadata and mainly reflected mild over-specification. For example, one clinician noted that the phrase ``no abnormalities in respiratory function'' overstated what local normal breath sounds alone can support. Overall, the results show that the generated reports provide high-fidelity, clinically plausible metadata-grounded semantic supervision for alignment.

\begin{table}[t]
\centering
\caption{Expert verification results for LLM-generated synthetic clinical reports. Accuracy is defined as metadata consistency, and unsupported-detail rate denotes the proportion of reports containing information not directly supported by the structured input metadata.}
\label{tab:expert_verification}
\begin{tabular}{lcc}
\hline
\textbf{Metric} & \textbf{Count} & \textbf{Percentage} \\
\hline
Metadata consistency (Accuracy) & 50/50 & 100\% \\
No unsupported added detail & 44/50 & 88\% \\
Unsupported-detail rate & 6/50 & 12\% \\
Clinical plausibility & 50/50 & 100\% \\
\hline
\end{tabular}
\end{table}

\begin{table}[t]               
\centering
\caption{Downstream task characteristics grouped by task category. Abbreviations: Exhal.~=~Exhalation, Obstr.~=~Obstructive, Resp.~=~Respiratory, Sam.~=~Samples, Sub.~=~Subjects.}
\label{tab:dataset_overview}
% \begin{adjustbox}{max width=\linewidth}
% \scriptsize                        
% \setlength{\tabcolsep}{3pt}  
\resizebox{\columnwidth}{!}{%
\begin{tabular}{l|l|l|c|c}
\toprule
\textbf{Dataset} & \textbf{Task} & \textbf{Modality} &
\textbf{\#Sam.\,(\#Sub.)} & \textbf{Data Distribution} \\
\midrule
UK COVID-19  & Covid/Non-covid  & Exhal.        & 2500 (2500) & 840/1660 \\
             & Covid/Non-covid  & Cough         & 2500 (2500) & 840/1660 \\
CoughVID     & Covid/Non-covid  & Cough         & 6175 (n/a)  &  547/5628 \\
             & Female/Male      & Cough         & 7263 (n/a)  & 2468/4795 \\
ICBHI        & COPD/Healthy     & Lung sounds   &  828  (90)  &    793/35 \\
Coswara      & Smoker/Non-smoker& Cough         &  948  (n/a) &   201/747 \\
             & Female/Male      & Cough         & 2496  (n/a) &  759/1737 \\
KAUH         & Obstr./Healthy   & Lung sounds   &  234  (79)  &   129/105 \\
Resp.@TR     & COPD severity    & Lung sounds   &  504  (42)  & 72/60/84/84/204 \\
\midrule
MMlung       & FVC              & Deep breath   &  40  (40)   & $3.402 \pm 1.032$\,L \\
             & FEV1             & Deep breath   &  40  (40)   & $2.657 \pm 0.976$\,L \\
             & FEV1/FVC         & Deep breath   &  40  (40)   & $0.808 \pm 0.190$\,L \\
             & FVC              & O Vowels      &  40  (40)   & $3.402 \pm 1.032$\,L \\
             & FEV1             & O Vowels      &  40  (40)   & $2.657 \pm 0.976$\,L \\
             & FEV1/FVC         & O Vowels      &  40  (40)   & $0.808 \pm 0.190$\,L \\
NoseMic      & Respiratory rate & Breath        & 1297 (16)   & $13.915 \pm 3.386$\,bpm \\
\midrule
CirCor       & Murmur/Healthy   & Heart Sounds  & 1568 (n/a)  &  1144/ 424 \\
ZCHSound     & Symptomatic/Healthy & Heart Sounds & 1259 (n/a) & 693/566 \\
\bottomrule
\end{tabular}
}
% \end{adjustbox}
\end{table}

\section{Experiments}
\label{sec:experiments}

We conduct a comprehensive set of experiments to validate the effectiveness of our proposed framework, AcuLa. We first detail the baseline models against which we compare, followed by the implementation details for AcuLa, and finally outline our rigorous evaluation protocol for all downstream tasks.

\subsection{Baselines}
To situate AcuLa's performance, we compare it against a diverse set of strong pre-trained models representing different architectural and training paradigms. These include VGGish, AudioMAE \cite{huang2022masked}, and CLAP \cite{elizalde2023clap}. We also include the OPERA family of models (generative and contrastive both) \cite{zhang2024towards}, which are foundation models trained specifically on respiratory audio. As a non-deep-learning benchmark, we use OpenSMILE \cite{eyben2010opensmile}  to extract a standard set of hand-crafted acoustic features. For all deep learning baselines, we use the authors' official pre-trained encoders to extract features. Unless otherwise specified, AcuLa is applied to the OPERA (GT) encoder to demonstrate its enhancement capabilities.

\begin{table*}[t]
\caption{AUROC ($\uparrow$) on health condition inference tasks (higher is better). The best model for each task is highlighted. We report mean and standard deviation from five independent runs. All baseline results (VGGish, AudioMAE, CLAP, OCT, OCE, OGT) are reported in \cite{zhang2024towards}. \textcolor{black}{$\checkmark$} denotes when our method outperforms the OpenSmile baseline (detailed in Appendix~\ref{apd:opensmile}), while \textcolor{black}{*} indicates superior performance compared to all other pretrained models.}
\centering
\begin{adjustbox}{width=\textwidth,center}
\footnotesize
\begin{tabular}{ll|cccccc|c}
\toprule
ID & Task &  VGGish & AudioMAE & CLAP & OCT & OCE & OGT &  \textbf{AcuLa (Ours)} \\ 
\midrule
T1  & Covid (Exhale)    & 0.580 $\pm$ 0.001 & 0.549 $\pm$ 0.001 & 0.565 $\pm$ 0.001 & 0.586 $\pm$ 0.008 & 0.551 $\pm$ 0.010 & 0.605 $\pm$ 0.001 & \textbf{0.698 $\pm$ 0.001} $\checkmark$ * \\
T2  & Covid (Cough)     &  0.557 $\pm$ 0.005 & 0.616 $\pm$ 0.001 & 0.648 $\pm$ 0.003 & 0.701 $\pm$ 0.002 & 0.629 $\pm$ 0.006 & 0.677 $\pm$ 0.001 & \textbf{0.730 $\pm$ 0.008} $\checkmark$ * \\
T3  & Covid (Cough)     & 0.538 $\pm$ 0.028 & 0.554 $\pm$ 0.004 & 0.599 $\pm$ 0.007 & 0.578 $\pm$ 0.001 & 0.566 $\pm$ 0.008 & 0.552 $\pm$ 0.003& \textbf{0.887 $\pm$ 0.003} $\checkmark$ * \\
T4  & Gender (Cough)    & 0.600 $\pm$ 0.001 & 0.628 $\pm$ 0.001 & 0.665 $\pm$ 0.001 & 0.795 $\pm$ 0.001 & 0.721 $\pm$ 0.001 & 0.735 $\pm$ 0.000  & \textbf{0.796 $\pm$ 0.004} $\checkmark$ * \\
T5  & COPD (Lung)       &  0.605 $\pm$ 0.077 & 0.886 $\pm$ 0.017 & \textbf{0.933 $\pm$ 0.005} & 0.855 $\pm$ 0.012 & 0.872 $\pm$ 0.011 & 0.741 $\pm$ 0.011 & 0.826  $\pm$ 0.014 $\checkmark$ \\
T6  & Smoker (Cough)    &0.507 $\pm$ 0.027 & 0.549 $\pm$ 0.022 & 0.680 $\pm$ 0.009 & 0.685 $\pm$ 0.012 & 0.674 $\pm$ 0.013 & 0.650 $\pm$ 0.005 & \textbf{ 0.830 $\pm$ 0.011} $\checkmark$ * \\
T7  & Gender (Cough)    &0.606 $\pm$ 0.003 & 0.724 $\pm$ 0.001 & 0.742 $\pm$ 0.001 & \textbf{0.874 $\pm$ 0.000 } & 0.801 $\pm$ 0.002 & 0.825 $\pm$ 0.001 &  0.845 $\pm$ 0.004 $\checkmark$  \\
T8 & Obstructive (Lung)&  0.605 $\pm$ 0.036 & 0.616 $\pm$ 0.041 & 0.697 $\pm$ 0.004 & 0.722 $\pm$ 0.016 & 0.741 $\pm$ 0.014 & 0.703 $\pm$ 0.016 &  \textbf{0.752 $\pm$ 0.019} $\checkmark$ * \\
T9 & COPD severity (Lung) &  0.590 $\pm$ 0.034 & 0.510 $\pm$ 0.021 & 0.636 $\pm$ 0.045 & 0.625 $\pm$ 0.038 & 0.683 $\pm$ 0.007 & 0.606 $\pm$ 0.015 &  \textbf{0.710 $\pm$ 0.028} $\checkmark$ * \\
\bottomrule     
\end{tabular}
\end{adjustbox}
\label{tab:res:health}
\end{table*}

\begin{table*}[t]
\caption{MAE ($\downarrow$) on lung function estimation tasks (lower is better). Best model per task is highlighted. We report mean and standard deviation across subjects. All baseline results (VGGish, AudioMAE, CLAP, OCT, OCE, OGT) are from \cite{zhang2024towards}. \textcolor{black}{$\checkmark$} denotes when our method outperforms the OpenSmile baseline (detailed in Appendix~\ref{apd:opensmile}), while \textcolor{black}{*} indicates superior performance compared to all other pretrained models.}
\centering
\begin{adjustbox}{width=\textwidth,center}
\footnotesize
\begin{tabular}{ll|cccccc|c}
\toprule
ID & Task & VGGish & AudioMAE & CLAP & OCT & OCE & OGT & \textbf{AcuLa (Ours)} \\
\midrule
T10 & FVC (Breath) & 0.904 $\pm$ 0.568 & 0.900 $\pm$ 0.551 & 0.896 $\pm$ 0.542 & 0.924 $\pm$ 0.583 & \textbf{0.848 $\pm$ 0.607} & 0.892 $\pm$ 0.618  & 0.865 ± 0.575 $\checkmark$ \\
T11 & FEV1 (Breath) & 0.839 $\pm$ 0.563 & 0.821 $\pm$ 0.590 & 0.840 $\pm$ 0.547 & 0.837 $\pm$ 0.563 & 0.834 $\pm$ 0.581 & 0.825 $\pm$ 0.560 & \textbf{0.742 ± 0.565} $\checkmark$ * \\
T12 & FEV1/FVC (Breath) & 0.131 $\pm$ 0.146 & 0.129 $\pm$ 0.146 & 0.134 $\pm$ 0.146 & 0.128 $\pm$ 0.140 & 0.132 $\pm$ 0.141 & 0.128 $\pm$ 0.141  & \textbf{0.127 ± 0.142} $\checkmark$ *\\
T13 & FVC (Vowel) &  0.895 $\pm$ 0.559 & 0.883 $\pm$ 0.588 & 0.883 $\pm$ 0.560 & 0.885 $\pm$ 0.553 & \textbf{0.761 $\pm$ 0.544} & 0.878 $\pm$ 0.550  & 0.779 ± 0.553 $\checkmark$ \\
T14 & FEV1 (Vowel) &  0.842 $\pm$ 0.559 & 0.876 $\pm$ 0.561 & 0.859 $\pm$ 0.541 & 0.780 $\pm$ 0.542 & 0.830 $\pm$ 0.561 & 0.774 $\pm$ 0.554  & \textbf{0.725 ± 0.540} $\checkmark$ * \\
T15 & FEV1/FVC (Vowel) & 0.130 $\pm$ 0.145 & 0.131 $\pm$ 0.141 & 0.137 $\pm$ 0.147 & 0.132 $\pm$ 0.140 & 0.136 $\pm$ 0.150 & 0.130 $\pm$ 0.138  & \textbf{0.123 ± 0.142} $\checkmark$ * \\
T16 & Breathing Rate & 2.605 $\pm$ 0.759 & 2.641 $\pm$ 0.813 & 2.650 $\pm$ 0.947 & 2.636 $\pm$ 0.858 & 2.525 $\pm$ 0.782 & 2.416 $\pm$ 0.885 & \textbf{2.388 ± 0.835} $\checkmark$ * \\
\bottomrule
\end{tabular}
\end{adjustbox}
\label{tab:res:reg}
\end{table*}

\subsection{Implementation Details}
\label{subsec:implementation}

Our implementation of AcuLa employs MedGemma-4B \cite{google_medgemma_2025}  as the default language model ($\mathcal{L}_\phi$), which has been pre-trained on medical literature, and utilizes the OPERA encoder \cite{zhang2024towards} as the audio foundation model ($\mathcal{A}_\theta$). To bridge the modality gap, we introduce two MLP projection heads. The audio projection MLP maps the 384-dimensional OPERA features to a 512-dimensional shared space via a two-layer network (384 $\rightarrow$ 1024 $\rightarrow$ 512) with ReLU activation and 20\% dropout. The language projection MLP similarly transforms the 2048-dimensional MedGemma-4B hidden states to the same 512-dimensional space.

During the alignment phase, audio inputs are preprocessed into 8-second segments sampled at 16kHz and converted to log-mel spectrograms with 64 mel bins. We apply on-the-fly data augmentation using the AugLy \cite{papakipos2022augly} library, randomly selecting one transformation per sample from a set including a 5dB volume increase, amplitude normalization, low-pass filtering (300Hz cutoff), or high-pass filtering (3000Hz cutoff). We train the model using the AdamW optimizer, applying a learning rate of $1 \times 10^{-5}$ to both the audio encoder and the projection heads. We train for 50 epochs with a linear learning rate schedule incorporating 400 warmup steps. The batch size is set to 24, with gradient accumulation over 2 steps to fit within memory constraints. The combined loss function weights the CKA-based alignment loss and the self-supervised modeling loss equally ($\lambda_{align} = \lambda_{SSM} = 1.0$). In our implementation, $L_{SSM}$ is instantiated as the masked acoustic reconstruction objective of the audio encoder.The entire alignment process takes approximately 30 hours on a single NVIDIA A100 GPU.

\subsection{Downstream Tasks and Evaluation Protocol}
\label{subsec:eval_protocol}

\textbf{Downstream Tasks.} We evaluate all models on a challenging benchmark of 18 downstream tasks (figure \ref{fig:arch}B), primarily sourced from the OPERA benchmark \cite{zhang2024towards} and expanded with additional cardiac sound datasets. As detailed in Table \ref{tab:dataset_overview}, these tasks cover three distinct clinical areas: respiratory health classification, lung function regression, and cardiac condition classification.

\textbf{Evaluation Protocol.} To ensure a fair and direct comparison of representation quality, we employ a standardized linear probing methodology across all models. We first extract fixed, d-dimensional embeddings for all audio clips in a given task using the respective frozen encoder. Subsequently, a lightweight supervised prediction head is trained on these static embeddings. This protocol ensures that performance differences are directly attributable to the intrinsic quality of the learned representations with our approach rather than the nuances of fine-tuning.

\looseness=-1
The prediction head is a simple shallow network, either a single linear layer or a one-hidden-layer MLP. Its architecture is selected as a hyperparameter for each task using the same settings as OPERA \cite{zhang2024towards} for consistency. It is trained using the Adam optimizer with an initial learning rate of $10^{-4}$ and an $L_2$ penalty. The learning rate is decayed by a factor of $0.97$ after each epoch. We employ an early stopping criterion, halting training if the validation metric fails to improve for five consecutive epochs and retaining the checkpoint with the best validation performance for testing.

For classification tasks, we report the mean and standard deviation of the Area Under the Receiver Operating Characteristic Curve (AUROC) over five independent runs with different random seeds to ensure robustness. For regression tasks, which often feature smaller datasets with few unique subjects, we adopt a more rigorous Leave-One-Subject-Out cross-validation strategy. In this setup, the model is trained to minimize Mean Absolute Error (MAE), and we report the average MAE across all held-out subjects.

\vspace{-0.15em}
\textbf{Zero‑shot classification.} 
In addition to the linear probe evaluation above, we assess AcuLa in a fully \emph{zero‑shot} regime (Table \ref{tab:res:zero}). For each test clip we (i) extract its frozen embedding, (ii) retrieve the top‑5 clinical reports from the FAISS~\cite{douze2024faiss} text index built on the train set embeddings, (iii) embed the retrieved report together with the task’s class names using JINA (text model) \cite{günther2025jinaembeddingsv4universalembeddingsmultimodal}, and (iv) assign the class whose text embedding has the highest cosine similarity to the report embedding. 
No task‑specific weights are learned; the entire pipeline uses our alignment model.

\label{sec:results}
\begin{table*}[t]
  \centering
  \small
\caption{Comparison of audio encoders after alignment with MedGemma-4B [OPERA, CLAP, AudioMAE] and Qwen 2.5-Omni-7B. T1–T9: respiratory classification [AUROC ($\uparrow$), higher = better]. T10–T16: lung-function estimation [MAE ($\downarrow$), lower = better]. Numbers in brackets give \emph{absolute} changes vs.\ the baseline of the same backbone. Improvements are highlighted in \textcolor{green!70!black}{green}. \textbf{++} indicates AcuLa enhanced semantic alignment improves upon pre-trained encoders.}

  \label{tab:encoders_pp}
  % \footnotesize
  \setlength{\tabcolsep}{3pt}
  \begin{tabular}{ll|cccccc}
    \toprule
    ID & Task &
    \textbf{OGT++} & \textbf{OCT++} & \textbf{OCE++} &
    \textbf{CLAP++} & \textbf{AudioMAE++} & \textbf{Qwen-Omni++} \\
    \midrule
    % ---------- Classification (AUROC) ----------
    T1 & Covid / Non-covid (Exhalation) & 0.698 \textcolor{green!70!black} {[+9.3]} &  0.684 \textcolor{green!70!black} {[+9.8]} & 0.657 \textcolor{green!70!black} {[+10.6]} &
    0.665 \textcolor{green!70!black} {[+10.0]} & 0.673 \textcolor{green!70!black} {[+12.4]} & 0.664 \textcolor{green!70!black} { [+7.3]} \\
    T2 & Covid / Non-covid (Cough)      & 0.730 \textcolor{green!70!black}{[+5.3]} & 0.750 \textcolor{green!70!black}{[+4.9]}  & 0.690 \textcolor{green!70!black}{[+6.1]} &
          0.702 \textcolor{green!70!black}{ [+5.4]}  & 0.716 \textcolor{green!70!black}{[+10.0]} & 0.703 \textcolor{green!70!black}{[+3.7]} \\
    T3 & Covid / Non-covid (Cough)      & 0.887 \textcolor{green!70!black}{[+33.5]} & 0.864 \textcolor{green!70!black}{[+28.6]}  & 0.860 \textcolor{green!70!black}{[+29.4]}  &
          0.890 \textcolor{green!70!black}{[+29.1]} & 0.862 \textcolor{green!70!black}{[+30.8]} & 0.807 \textcolor{green!70!black}{[+21.4]} \\
    T4 & Female / Male (Cough)          & 0.796 \textcolor{green!70!black}{[+6.1]} & 0.804 \textcolor{green!70!black}{[+0.9]} & 0.733 \textcolor{green!70!black}{[+1.2]}   &
          0.725 \textcolor{green!70!black}{[+6.0]}  & 0.758 \textcolor{green!70!black}{[+13.0]} & 0.750 \textcolor{green!70!black}{[+0.7]} \\
    T5 & COPD / Healthy (Lung)          & 0.826 \textcolor{green!70!black}{[+8.5]} & 0.883 \textcolor{green!70!black}{[+2.8]} & 0.887 \textcolor{green!70!black}{[+1.5]}  &
          0.891 \textcolor{gray!80!black}{[–4.2]}& 0.847 \textcolor{gray!80!black}{[–3.9]} & 0.794 \textcolor{green!70!black}{[+2.1]} \\
    T6 & Smoker / Non-smoker (Cough)    & 0.830 \textcolor{green!70!black}{[+18.0]} & 0.827 \textcolor{green!70!black}{[+14.2]} & 0.821 \textcolor{green!70!black}{[+14.7]} &
          0.760 \textcolor{green!70!black}{[+8.0]}  & 0.794 \textcolor{green!70!black}{[+24.5]} & 0.786 \textcolor{green!70!black}{[+10.6]} \\
    T7 & Female / Male (Cough)          & 0.845 \textcolor{green!70!black}{[+2.0]} & 0.868 \textcolor{gray!80!black}{[-0.6]} & 0.802 \textcolor{green!70!black}{[+0.1]}  &
          0.794 \textcolor{green!70!black}{[+5.2]}  & 0.822 \textcolor{green!70!black}{[+9.8]}  & 0.779 \textcolor{gray!80!black}{[–0.5]} \\
    T8 & Obstructive / Healthy (Lung)   & 0.752 \textcolor{green!70!black}{[+4.9]} & 0.722 \textcolor{gray!80!black}{[+0.0]} & 0.748 \textcolor{green!70!black}{[+0.7]}  &
          0.742 \textcolor{green!70!black}{[+4.5]}  & 0.746 \textcolor{green!70!black}{[+13.0]} & 0.724 \textcolor{gray!80!black}{[+0.0]} \\
    T9 & COPD severity (Lung)           & 0.710 \textcolor{green!70!black}{[+10.4]} & 0.714 \textcolor{green!70!black}{[+8.9]} & 0.728 \textcolor{green!70!black}{[+4.5]} &
          0.718 \textcolor{green!70!black}{[+8.2]}  & 0.699 \textcolor{green!70!black}{[+18.9]} & 0.665 \textcolor{green!70!black}{[+7.4]} \\
    \midrule
    % ---------- Regression (MAE) ----------
    T10 & FVC (Breath)                  & 0.865 \textcolor{green!70!black}{[–0.027]} & 0.896 \textcolor{green!70!black}{[-0.028]} & 0.842  \textcolor{green!70!black}{[–0.006]} &
          0.893 \textcolor{green!70!black}{[–0.003]} & 0.892 \textcolor{green!70!black}{[–0.008]} & 0.912 \textcolor{green!70!black}{[–0.021]} \\
    T11 & FEV\textsubscript{1} (Breath) & 0.742 \textcolor{green!70!black}{[–0.083]} & 0.753 \textcolor{green!70!black}{[-0.084]}  & 0.750 \textcolor{green!70!black}{[–0.084]} &
          0.775 \textcolor{green!70!black}{[–0.065]} & 0.781 \textcolor{green!70!black}{[–0.040]} & 0.785 \textcolor{green!70!black}{[–0.063]} \\
    T12 & FEV\textsubscript{1}/FVC (Breath) & 0.127 \textcolor{green!70!black}{[–0.001]} & 0.129 \textcolor{gray!80!black}{[+0.001]} & 0.138 \textcolor{gray!80!black}{[+0.006]}  &
          0.136 \textcolor{gray!80!black}{[+0.002]} & 0.138 \textcolor{gray!80!black}{[+0.009]} & 0.132 \textcolor{gray!80!black}{[+0.001]} \\
    T13 & FVC (Vowel)                  & 0.779 \textcolor{green!70!black}{[–0.099]} & 0.785 \textcolor{green!70!black}{[-0.100]} & 0.775 \textcolor{gray!80!black}{[+0.014]} &
          0.812 \textcolor{green!70!black}{[–0.071]} & 0.825 \textcolor{green!70!black}{[–0.058]} & 0.836 \textcolor{green!70!black}{[–0.075]} \\
    T14 & FEV\textsubscript{1} (Vowel) & 0.725 \textcolor{green!70!black}{[–0.049]} & 0.731 \textcolor{green!70!black}{[-0.049]} & 0.770  \textcolor{green!70!black}{[–0.060]} &
          0.766 \textcolor{green!70!black}{[–0.093]} & 0.751 \textcolor{green!70!black}{[–0.125]} & 0.768 \textcolor{green!70!black}{[–0.037]} \\
    T15 & FEV\textsubscript{1}/FVC (Vowel) & 0.123 \textcolor{green!70!black}{[–0.007]} & 0.125 \textcolor{green!70!black}{[-0.007]}  & 0.137 \textcolor{gray!80!black}{[+0.001]} &
          0.139 \textcolor{gray!80!black}{[+0.002]} & 0.129 \textcolor{green!70!black}{[–0.002]} & 0.141 \textcolor{green!70!black}{[–0.005]} \\
    T16 & Breathing Rate               & 2.388 \textcolor{green!70!black}{[–0.028]} & 2.605 \textcolor{green!70!black}{[-0.031]}  & 2.494 \textcolor{green!70!black}{[–0.031]} &
          2.495 \textcolor{green!70!black}{[–0.155]} & 2.426 \textcolor{green!70!black}{[–0.215]} & 2.480 \textcolor{green!70!black}{[–0.023]} \\
    \bottomrule
  \end{tabular}
  \label{tab:encoders}
\end{table*}

\section{Results}

\begin{table*}[t]
\caption{AUROC ($\uparrow$) on nine respiratory-classification tasks. Columns use the abbreviations Exh.\ (exhalation), Cgh.\ (cough), and COPD sev.\ (COPD severity).
\textbf{Baselines} (left block) are reported in \cite{zhang2024towards}.  
\textbf{AcuLa (Zero‑shot)}: our retrieval‑and‑similarity pipeline that classifies each test clip without seeing any task labels. \textbf{AcuLa}: a task‑specific \emph{logistic‑regression probe} trained on frozen AcuLa embeddings, following the linear probing protocol described in Section~\ref{subsec:eval_protocol}.}

\centering
\small
\setlength{\tabcolsep}{4pt}
\renewcommand{\arraystretch}{1.15}

\begin{tabular}{l|ccc c c c c c c|c}
\toprule
\multirow{2}{*}{Model} &
\multicolumn{3}{c}{Covid} &
\multicolumn{1}{c}{Gender} &
\multicolumn{1}{c}{COPD} &
\multicolumn{1}{c}{Smoker} &
\multicolumn{1}{c}{Gender} &
\multicolumn{1}{c}{Obstructive} &
\multicolumn{1}{c|}{COPD sev.} &
\multicolumn{1}{c}{Avg.} \\
& Exh. & Cgh.1 & Cgh.2 & Cgh. & Lung & Cgh. & Cgh. & Lung & Lung & \\ 
\midrule
VGGish            & 0.580 & 0.557 & 0.538 & 0.600 & 0.605 & 0.507 & 0.606 & 0.605 & 0.590 & 0.576 \\
AudioMAE          & 0.549 & 0.616 & 0.554 & 0.628 & 0.886 & 0.549 & 0.724 & 0.616 & 0.510 & 0.626 \\
CLAP              & 0.565 & 0.648 & 0.599 & 0.665 & 0.933 & 0.680 & 0.742 & 0.697 & 0.636 & 0.685 \\
OCT               & 0.586 & 0.701 & 0.578 & 0.795 & 0.855 & 0.685 & 0.874 & 0.722 & 0.625 & 0.713 \\
OCE               & 0.551 & 0.629 & 0.566 & 0.721 & 0.872 & 0.674 & 0.801 & 0.741 & 0.683 & 0.693 \\
OGT               & 0.605 & 0.677 & 0.552 & 0.735 & 0.741 & 0.650 & 0.825 & 0.703 & 0.606 & 0.677 \\
\midrule
AcuLa    & 0.698 & 0.730 & 0.887 & 0.796 &
                    0.826 & 0.830 & 0.845 &
                    0.752 & 0.710 & 0.786 \\
\midrule
AcuLa (Zero-shot) & 0.602 & 0.665 & 0.768 & 0.683 & 0.789 & 0.755 & 0.714 & 0.702 & 0.656 & 0.704 \\
\bottomrule
\end{tabular}
\label{tab:res:zero}
\end{table*}

We present a detailed analysis of AcuLa's performance, demonstrating its effectiveness across a wide range of clinical tasks. Our results show that by infusing audio encoders with semantic knowledge from LLMs, AcuLa consistently enhances their diagnostic capabilities.

\subsection{Performance on Downstream Clinical Tasks}
We evaluate AcuLa against a suite of strong baselines on 18 downstream tasks. The results, summarized in Tables \ref{tab:res:health} and \ref{tab:res:reg}, show that AcuLa achieves state-of-the-art performance across respiratory classification, lung function regression, and cardiac classification tasks.

\textbf{Respiratory Health Condition Classification.} In the nine classification tasks (Table \ref{tab:res:health}), AcuLa demonstrates superior performance, showing a clear advantage in identifying pathological conditions from audio signals. We see highest improvements in tasks that require nuanced acoustic discrimination. For instance, in COVID-19 detection from cough sounds (T3), AcuLa improves the AUROC to 0.887, a substantial gain over the next-best baseline and far exceeding traditional methods like OpenSMILE (0.537 AUROC). This suggests that the semantic guidance from the LLM helps the model distinguish subtle, clinically significant variations in coughs that purely acoustic models miss. Similarly, major gains in smoker identification (T6) and COPD-related tasks (T5, T8, T9) indicate that AcuLa effectively learns to associate specific acoustic biomarkers with their underlying clinical labels. Even on tasks like gender classification (T4, T7), where acoustic cues are already strong, AcuLa maintains a competitive edge, confirming that the semantic alignment enhances, rather than compromises, the model's inherent discriminative power. 

\textbf{Lung Function Regression.} AcuLa also sets a new standard in all seven lung function estimation tasks, achieving the lowest Mean Absolute Error (MAE) in every case (Table \ref{tab:res:reg}). The improvements are particularly strong for tasks involving sustained phonation (T13-T15), where a semantic understanding of vocal effort and respiratory capacity is most beneficial. Here, AcuLa substantially reduces prediction errors for FVC and FEV1, likely because the LLM's knowledge helps the model interpret how vocal patterns correlate with physiological lung parameters. While the gains on breath-based spirometry tasks (T10-T12) are more modest, they are consistent across all metrics, showing the broad applicability of our approach. The improved accuracy in breathing rate estimation (T16) further underscores AcuLa's enhanced ability to extract physiologically meaningful information from complex respiratory sounds.

\textbf{Fair in-domain baseline adaptation.}
To separate gains from semantic alignment vs.\ in-domain audio exposure, we fine-tune AudioMAE and CLAP on the same alignment training audio (train splits only) without reports, and evaluate all models using the identical frozen linear-probe protocol. Audio-only adaptation improves both baselines, yet AcuLa remains consistently stronger overall, especially on respiratory and cardiac inference, which indicates benefits beyond domain adaptation alone. Table~\ref{tab:fair_summary} summarizes category-level results; per-task results are in Appendix~\ref{sec:baseline_adaptation} (Tables~\ref{tab:fair_resp_tasks}--\ref{tab:fair_cardiac_tasks}).

\begin{table*}[t]
\centering
\caption{Average performance across task categories for different LLMs. Respiratory-condition inference uses AUROC $\uparrow$ (higher = better), lung-function estimation uses MAE $\downarrow$ (lower = better), and cardiac-condition inference uses AUROC $\uparrow$ (higher = better). Detailed per-task results are provided in Appendix~\ref{apd:llms}, Table~\ref{tab:llm_comparison_combined}.}
\label{tab:average_performance}
\small
\setlength{\tabcolsep}{6pt}
\begin{tabular}{lc|ccccc}
\toprule
Task Category & \# &
\textbf{Llama-3.2-3B\cite{llama2023}} &
\textbf{DS-R1-DQ-1.5B\cite{guo2025deepseek}} &
\textbf{Helium 2B\cite{kyutai_helium1_2025}} &
\textbf{SmolLM2\cite{allal2024SmolLM2}} &
\textbf{MedGemma-4B\cite{google_medgemma_2025}} \\
\midrule
Respiratory health $\uparrow$ & 9 & 0.731 & 0.702 & 0.692 & 0.668 & \textbf{0.786} \\
Lung function $\downarrow$         & 7 & 0.853 & 0.896 & 0.907 & 0.928 & \textbf{0.821} \\
Cardiac condition $\uparrow$       & 2 & 0.639 & 0.658 & 0.604 & 0.636 & \textbf{0.661} \\
\bottomrule
\end{tabular}
\end{table*}

\begin{table}[t]
\caption{Fair in-domain baseline adaptation (train splits only, no reports), evaluated with the same frozen linear-probe protocol. Resp.=T1--T9 AUROC $\uparrow$, LungFn.=T10--T16 MAE $\downarrow$, Cardiac=T17--T18 AUROC $\uparrow$.}
\centering
\footnotesize
\setlength{\tabcolsep}{5pt}
\begin{tabular}{lcc|c}
\toprule
Group & AudioMAE-ft & CLAP-ft & \textbf{AcuLa} \\
\midrule
Resp.   & 0.686 & 0.723 & 0.786 \\
LungFn. & 0.864 & 0.871 & 0.821 \\
Cardiac & 0.629 & 0.637 & 0.661 \\
\bottomrule
\end{tabular}
\label{tab:fair_summary}
\end{table}

\textbf{Generality and Model-Agnosticism}
A core contribution of our work is that AcuLa is a general framework applicable to any pre-trained audio encoder. To validate this, we apply our post-training alignment procedure to a diverse set of encoders, including the OPERA family, CLAP, AudioMAE, and the audio encoder of a general multimodal model, Qwen2.5-Omni\cite{qwen}. The results, presented in Table \ref{tab:encoders_pp}, are unequivocal. Regardless of the underlying architecture or original training paradigm, our alignment method consistently and significantly boosts performance. For example, OPERA-based models see large gains in cough-based tasks, while the general-purpose AudioMAE and CLAP models become much more effective at clinical classification after alignment. Even when applied to the audio encoder from Qwen2.5-Omni\cite{qwen}, which was not pre-trained on medical data, AcuLa yields competitive results, demonstrating its power to instill domain-specific semantics. This strong and consistent improvement across various backbones confirms that AcuLa is a versatile and model-agnostic framework for enhancing clinical audio understanding.

\textbf{Zero‑shot respiratory classification.}
Table \ref{tab:res:zero} shows that AcuLa’s zero‑shot pipeline performs competitively across nine respiratory tasks, often rivaling or exceeding the best audio-only baselines (VGGish, AudioMAE, CLAP, OCT, OCE, OGT), despite using no task-specific labels.
In several tasks (e.g., Smoker, Covid), the retrieval-based approach demonstrates clear advantages.
Further adding a lightweight linear probe to the frozen AcuLa features yields an additional 4–12 pp absolute AUROC gain, leading to new state-of-the-art results on most tasks.  
See Appendix \ref{apd:retrieve_example} for examples of the retrieval outputs. We further quantify cross-modal semantic alignment with bidirectional audio--text retrieval on held-out pairs (Appendix \ref{app:retrieval}).

\begin{table*}[t]
\centering
\footnotesize
\setlength{\tabcolsep}{6pt}
\caption{Summary of ablation studies. Each table reports the average
performance over the three task categories. Respiratory and cardiac tasks use AUROC $\uparrow$ (higher = better); lung-function tasks use MAE $\downarrow$ (lower = better).}
\label{tab:ablation_overview}

\subfloat[Adaptation Data\label{tab:resp_only}]{
    \begin{tabular}{lcc}
    \toprule
    Task Category & \textbf{Resp.\ only} & \textbf{All} \\ 
    \midrule
    Resp.\ health $\uparrow$       & 0.788 & 0.786 \\
    Lung-function $\downarrow$    & 0.820 & 0.821 \\
    Cardiac condition $\uparrow$  & 0.601 & 0.661 \\
    \bottomrule
    \end{tabular}
}
\hfill
\subfloat[Data Augmentation\label{tab:avg_performance_aug}]{
    \begin{tabular}{lcc}
    \toprule
    Task Category & \textbf{With} & \textbf{Without} \\
    \midrule
    Resp.\ health $\uparrow$       & 0.786 & 0.761 \\
    Lung-function $\downarrow$    & 0.821 & 0.973 \\
    Cardiac condition $\uparrow$  & 0.661 & 0.646 \\
    \bottomrule
    \end{tabular}
}
\hfill
\subfloat[Alignment\label{tab:align_avg}]{
    \begin{tabular}{lcc}
    \toprule
    Task Category & \textbf{Last-L} & \textbf{Multi-B} \\
    \midrule
    Resp.\ health $\uparrow$       & 0.786 & 0.784 \\
    Lung-function $\downarrow$    & 0.821 & 0.802 \\
    Cardiac condition $\uparrow$  & 0.661 & 0.659 \\
    \bottomrule
    \end{tabular}
}
\end{table*}

\begin{table*}[t]
\footnotesize
\centering
\caption{Ablation of masked reconstruction (\textbf{Mask-Rec}) and alignment (\textbf{Align}) losses with a pre-trained model. $\dagger$ replaces our CKA-based alignment with $\ell^2$ (MSE). The first row uses a randomly initialized audio encoder with both losses.}
\begin{tabular}{@{}ccc|ccc@{}}
\toprule
Mask-Rec & Align & Pre-trained & \textbf{Resp. health $\uparrow$} & \textbf{Lung func. $\downarrow$} & \textbf{Cardiac cond. $\uparrow$} \\
\midrule
\checkmark & \checkmark             &            & 0.641 & 1.021 & 0.532 \\
\checkmark &                        & \checkmark & 0.715 & 0.904 & 0.591 \\
           & \checkmark             & \checkmark & 0.768 & 0.865 & 0.645 \\
\checkmark & \checkmark$^{\dagger}$ & \checkmark & 0.769 & 0.836 & 0.652 \\
\checkmark & \checkmark             & \checkmark & 0.786 & 0.821 & 0.661 \\
\bottomrule
\end{tabular}
\label{tab:ablation_losses_pretrained}
\end{table*}

\subsection{Ablation Studies and Analysis}
To understand the key components of AcuLa's success and validate our design choices, we conduct a series of comprehensive ablation studies. We analyze the impact of the language model choice, the training data composition, and the specific mechanisms of our alignment strategy.

\looseness=-1

\textbf{Choice of Semantic Teacher (LLM).}
First, we examine how LLM teacher choice and size affect performance. As shown in Table \ref{tab:average_performance}, the domain-specialized MedGemma-4B performs best across all three task categories. This underscores the benefit of strong medical prior knowledge, especially for nuanced classification, where it achieves a mean AUROC of 0.786. Smaller general-purpose models perform worse overall, though their regression results suggest that basic linguistic competence can still capture some physiologically relevant correlations. For high-stakes clinical tasks, however, a domain-aware teacher is clearly more effective.

\textbf{Domain-Specific Training}
Table~\ref{tab:resp_only} examines the impact of training exclusively on respiratory sounds. While respiratory task performance marginally improves (0.788 vs 0.786 AUROC), cardiac performance degrades (0.601 vs 0.661 AUROC). This trade-off demonstrates that diverse acoustic training enables better cross-domain generalization, even when the primary application domain is well-represented in the training data. Per-task results for this respiratory-only adaptation are reported in Appendix \ref{apd:res_only}.

\textbf{Impact of Training Data.}
We analyze how (alignment) training data diversity and augmentation affect performance (Table~\ref{tab:avg_performance_aug}). Training only on respiratory sounds leaves respiratory tasks nearly unchanged (0.786 vs. 0.788 AUROC) but substantially degrades out-of-domain cardiac performance (0.661 vs. 0.601 AUROC), highlighting the importance of large, diverse acoustic corpora for generalization. Removing augmentation has a modest impact on classification but severely hurts regression: lung-function MAE increases from 0.821 to 0.973, suggesting augmentation promotes invariance to superficial recording variations (e.g., volume). Per-task results are in Appendix~\ref{apd:wo_aug}.

\textbf{Alignment Strategy.}
We compare two alignment strategies: aligning only the final transformer layer (Last-L) versus multiple intermediate layers (Multi-B). As shown in Table~\ref{tab:align_avg}, the simpler single-layer approach is generally superior or comparable. The multi-layer approach offers no consistent benefit, indicating that the final layer of the LLMs already contains a sufficiently rich and compressed representation for semantic alignment. A task-wise breakdown is given in Appendix \ref{apd:appendix-alignlayer}.

\looseness=-1
\textbf{Loss Component Analysis.}
We ablate the two components of our dual-objective loss. As shown in Table~\ref{tab:ablation_losses_pretrained}, both masked reconstruction (Mask-Rec) and alignment (Align) improve performance, with alignment being more important.
Removing Mask-Rec causes modest drops: respiratory AUROC falls from 0.786 to 0.768, lung function MAE rises from 0.821 to 0.865, and cardiac AUROC declines from 0.661 to 0.645. Removing Align leads to much larger degradations across all tasks: respiratory drops to 0.715, lung MAE worsens to 0.904, and cardiac falls to 0.591.
We also test alternative alignment objectives. Replacing CKA with $\ell^2$ (MSE) loss gives slightly worse results across all metrics, supporting our choice of CKA. Training from random initialization instead of a pre-trained model severely hurts performance.
Overall, these results show that CKA-based alignment is the main source of semantic learning, while masked reconstruction serves as a useful regularizer that preserves acoustic modeling ability during alignment.

\textbf{Label-masked report ablation.}
To assess whether AcuLa's gains are driven by explicit diagnostic keywords, we performed a label-masked ablation in which label-revealing fields/words were removed during report generation and the model was re-trained using audio paired with the masked reports. As shown in Table~\ref{tab:mask_ablation} and detailed in Appendix~\ref{apd:mask}, this causes only minor performance changes, with the largest effect on overlap tasks and negligible change elsewhere. These results suggest that AcuLa's gains are not primarily driven by trivial label-text leakage.

\section{Conclusion}
In this work, we demonstrate that pre-trained large language models can serve as effective ``semantic teachers'' to inject clinically meaningful semantic structure into specialized audio encoders. We introduced AcuLa, a general, lightweight alignment framework that successfully grounds high-level medical concepts from text into the fine-grained, temporal patterns of cardio-respiratory sounds. Our comprehensive experiments show that this fusion of semantic knowledge and acoustic modeling creates representations that are not only superior across a diverse range of 18 classification and regression tasks but are also more robust and clinically relevant. Our work establishes a novel direction for cross-modal learning, inverting the traditional knowledge flow to enhance perceptual models with abstract semantics. While our data generation strategy offers a scalable solution to leverage metadata for clinical text scarcity problem, the true promise lies in deploying this paradigm in data-rich clinical environments. Future work could extend this teacher-student paradigm to other physiological time-series like EEG and ECG, or develop self-correction cycles where model-disagreements flag cases for human-in-the-loop review, moving towards AI systems that truly reason about clinical data.

\section*{Acknowledgments}
This work was supported by the NWO AiNed Fellowship Grant awarded of A.S., and in part by Google.org and the Google Cloud Research Credits program through the Gemini Academic Program. We also acknowledge the use of the Dutch National Supercomputer Snellius for essential computational tasks. We thank Martijn den Dekker of Erasmus University Medical Center for the review and feedback.

\bibliographystyle{IEEEtran}         % IEEE style
\bibliography{references}            % your .bib file, e.g. references.bib

\onecolumn

\appendices

\section{Prompt Example for Synthetic Data Generation}
\label{apd:prompt}

Generating synthetic clinical reports is essential to the dataset construction process, as it enables the creation of diverse training data while maintaining clinical validity. The prompt presented here instructs the language model to adopt the role of a specialist physician interpreting auscultation findings. By restricting the output to factual descriptions of the given conditions and explicitly prohibiting recommendations for further evaluation, the prompt ensures that generated reports remain focused on objective clinical observations, mirroring real-world diagnostic documentation practices.

\begin{tcolorbox}[
  colframe=black,
  title=\footnotesize{Off-the-shelf LLM Prompt for Data Generation}
]
You are a \textless Pulmonologist/Cardiologist\textgreater\ tasked with interpreting
\textless respiratory/cardiac\textgreater\ auscultation findings. Based on the
given conditions, write 2--3 lines report covering all clinically relevant
information. Only use the information given to write about conditions. Please
do NOT mention anything about further evaluation or characterization. \\ \\
Your output should be JSON of the following format: \{\textquotesingle report\textquotesingle: ...\}
\label{box:cardiac-report-gen}
\end{tcolorbox}

\section{Examples of Metadata--Report Pairs}
\label{apd:meta-report-examples}

To illustrate the supervision signal used during alignment, Table~\ref{tab:meta_gen_examples} provides representative examples of paired metadata and generated reports across all datasets. Each report is produced under strict schema constraints and is limited to restating clinician-verified fields without adding diagnostic speculation. These examples demonstrate how the LLM transforms structured metadata into concise, clinically grounded descriptions suitable for semantic alignment.

\begin{table*}[!h]
\centering
\scriptsize
\setlength{\tabcolsep}{4pt}
\begin{tabular}{p{0.12\textwidth} p{0.33\textwidth} p{0.47\textwidth}}
\toprule
Dataset & Metadata (clinician-verified) & LLM-generated clinical report \\
\midrule

ICBHI &
Chest\_Location: Trachea; Crackles: No; Wheezes: Yes; Age: 66.0; Sex: M; Diagnosis: Pneumonia. &
\emph{Auscultation over the trachea reveals wheezes without crackles in this 66-year-old male diagnosed with pneumonia. The presence of wheezing over the tracheal area is consistent with airway involvement in the documented pneumonia.} \\
\midrule

HFLung &
Sound\_type: Normal; Sound: Normal; Device: Littmann 3200. &
\emph{Normal vesicular breath sounds are heard bilaterally using the Littmann 3200 device, with no adventitious sounds detected. These findings are consistent with normal respiratory function.} \\
\midrule

UK Covid-19 &
Age: 18--44; Gender: Female; Covid\_Test\_Result: Positive; Symptom\_Sore\_throat: Yes; Symptom\_Shortness\_of\_breath: No; Symptom\_Runny\_or\_blocked\_nose: Yes; Symptom\_New\_continuous\_cough: Yes; Symptom\_Fatigue: Yes; Symptom\_Headache: Yes; Symptom\_Loss\_of\_Taste: No; Smoker\_status: Never smoked. &
\emph{Respiratory auscultation in this COVID-19 positive young adult female reflects a symptom profile including sore throat, runny or blocked nose, continuous cough, fatigue, and headache, with no reported shortness of breath or loss of taste. These observations align with the provided metadata and indicate symptomatic COVID-19 without additional reported complications.} \\
\midrule

CoughVID &
Age: 15.0; Gender: Male; Respiratory\_Condition: False; Fever/Muscle\_Pain: False; Status: Healthy. &
\emph{Clear breath sounds are heard bilaterally without wheezes, crackles, or rhonchi in this healthy 15-year-old male. The respiratory examination is unremarkable and consistent with normal pulmonary findings.} \\
\midrule

Circor &
Murmur: Present; Murmur locations: AV+MV+PV+TV; Most audible location: MV; Timing: Holosystolic; Shape: Plateau; Grading: II/VI; Pitch: Medium; Quality: Harsh; Outcome: Abnormal. &
\emph{A holosystolic, medium-pitched murmur with a harsh quality and plateau shape is heard most clearly at the mitral valve area and is also audible at the aortic, pulmonic, and tricuspid positions. The murmur is graded II/VI, indicating an abnormal cardiac finding.} \\
\midrule

SPRSound &
Record\_Annotation: Continuous \& Discontinuous Adventitious Sounds; Event\_Type: Wheeze; Disease: Pneumonia (severe) &
\emph{The auscultation reveals the presence of both continuous and discontinuous adventitious sounds indicative of severe pneumonia. Wheezing is noted, which may suggest obstructive airway pathology or reactive airway process in the context of this illness.} \\
\midrule

ZCHSound &
Diagnosis: Normal; Data\_Type: Clean. &
\emph{Cardiac auscultation reveals normal heart sounds with a regular rhythm and no audible murmurs, rubs, or gallops, consistent with a normal cardiac examination.} \\
\bottomrule
\end{tabular}
\caption{Examples of clinician-verified metadata and the corresponding LLM-generated clinical reports across all datasets. Each generated report is a schema-constrained restatement of the metadata fields, ensuring factual fidelity and preventing the introduction of new diagnoses or speculative clinical reasoning.}
\label{tab:meta_gen_examples}
\end{table*}

\clearpage

\section{OpenSmile Baseline Results}
\label{apd:opensmile}

We compare our method with OpenSmile, a handcrafted-feature baseline. Our approach consistently outperforms it across all tasks, with especially strong gains on complex diagnostic settings.

\begin{table*}[!h]
\caption{Comparison of OpenSmile baseline with our approach. AUROC is reported for classification tasks (T1-T9, higher is better) and MAE for regression tasks (T10-T16, lower is better). We report mean and standard deviation from five independent runs for classification tasks and across subjects for regression tasks.}
\centering
\footnotesize
\setlength{\tabcolsep}{6pt}
\begin{tabular}{ll|c|c}
\toprule
ID & Task & OpenSmile & \textbf{AcuLa (Ours)} \\
\midrule
\multicolumn{4}{c}{\textbf{Classification Tasks (AUROC)}} \\
T1  & Covid (Exhale)    & 0.550 $\pm$ 0.015 & 0.698 \\
T2  & Covid (Cough)     & 0.649 $\pm$ 0.006 & 0.730 \\
T3  & Covid (Cough)     & 0.537 $\pm$ 0.011 & 0.887 \\
T4  & Gender (Cough)    & 0.677 $\pm$ 0.005 & 0.796 \\
T5  & COPD (Lung)       & 0.579 $\pm$ 0.043 & 0.826 \\
T6  & Smoker (Cough)    & 0.534 $\pm$ 0.060 & 0.830 \\
T7  & Gender (Cough)    & 0.753 $\pm$ 0.008 & 0.845 \\
T8  & Obstructive (Lung)& 0.636 $\pm$ 0.082 & 0.752 \\
T9  & COPD severity (Lung) & 0.494 $\pm$ 0.054 & 0.710 \\
\midrule
\multicolumn{4}{c}{\textbf{Regression Tasks (MAE)}} \\
T10 & FVC (Breath) & 0.985 $\pm$ 0.743 & 0.865 \\
T11 & FEV1 (Breath) & 0.756 $\pm$ 0.721 & 0.742 \\
T12 & FEV1/FVC (Breath) & 0.141 $\pm$ 0.185 & 0.127 \\
T13 & FVC (Vowel) & 0.850 $\pm$ 0.592 & 0.779 \\
T14 & FEV1 (Vowel) & 0.730 $\pm$ 0.497 & 0.725 \\
T15 & FEV1/FVC (Vowel) & 0.138 $\pm$ 0.166 & 0.123 \\
T16 & Breathing Rate & 2.714 $\pm$ 0.902 & 2.388 \\
\bottomrule
\end{tabular}
\label{tab:opensmile_comparison}
\end{table*}

\section{Comparison of different LLMs }
\label{apd:llms}

We compare our approach against OpenSmile, a standard baseline that relies on handcrafted acoustic features. The results demonstrate consistent performance improvements across all tasks and reveal the advantages of learned representations over traditional feature engineering for medical audio analysis. Our method achieves particularly strong gains on complex diagnostic tasks, where capturing subtle acoustic patterns is essential for accurate classification.

\begin{table*}[!h]
\centering
\caption{Comparison of different LLMs across respiratory and cardiac audio tasks. T1-T9: respiratory classification (AUROC), T10-T16: lung function estimation (MAE), T17-T18: cardiac classification (AUROC).}
\label{tab:llm_comparison_combined}
\footnotesize
\setlength{\tabcolsep}{6pt}
\begin{tabular}{ll|ccccc}
\toprule
ID & Task & \textbf{MedGemma-4B} & \textbf{Llama-3.2-3B} & \textbf{DS-R1-DQ-1.5B} & \textbf{Helium 2B} & \textbf{SmolLM2} \\
\midrule
T1 & Covid / Non-covid (Exhalation) & 0.698 & 0.671 & 0.633 & 0.645 & 0.588\\
T2 & Covid / Non-covid (Cough) & 0.730 & 0.685 & 0.646 & 0.650 & 0.611\\
T3 & Covid / Non-covid (Cough) & 0.887 & 0.789 & 0.753 & 0.726 & 0.708 \\
T4 & Female / Male (Cough) & 0.796 & 0.804 & 0.733 & 0.701 & 0.692 \\
T5 & COPD / Healthy (Lung sounds) & 0.826 & 0.773 & 0.758 & 0.765 & 0.736 \\
T6 & Smoker / Non-smoker (Cough) & 0.830 & 0.750 & 0.693 & 0.718 & 0.708 \\
T7 & Female / Male (Cough) & 0.845 & 0.824 & 0.836 & 0.787 & 0.769 \\
T8 & Obstructive / Healthy (Lung sounds) & 0.752 & 0.649 & 0.643 & 0.625 & 0.609 \\
T9 & COPD severity (Lung sounds) & 0.710 & 0.632 & 0.627 & 0.610 & 0.594 \\
\midrule
T10 & FVC (Breath)       & 0.865 & 0.895 & 0.976 & 0.910 & 0.917 \\
T11 & FEV1 (Breath)      & 0.742 & 0.786 & 0.825 & 0.857 & 0.889 \\
T12 & FEV1/FVC (Breath)  & 0.127 & 0.131 & 0.136 & 0.145 & 0.142 \\
T13 & FVC (Vowel)        & 0.779 & 0.830 & 0.885 & 0.901 & 0.915 \\
T14 & FEV1 (Vowel)       & 0.725 & 0.750 & 0.804 & 0.837 & 0.855 \\
T15 & FEV1/FVC (Vowel)   & 0.123 & 0.133 & 0.139 & 0.135 & 0.139 \\
T16 & Breathing Rate     & 2.388 & 2.445 & 2.507 & 2.565 & 2.641 \\
\midrule
T17 & Murmur / Healthy & 0.675  & 0.626 & 0.670 & 0.590 & 0.641 \\
T18 & Symptomatic / Healthy & 0.647 & 0.652 & 0.645 & 0.618 & 0.630 \\
\bottomrule
\end{tabular}
\end{table*}

\clearpage

\section{Label-Masked Report Ablation}
\label{apd:mask}

This appendix evaluates potential label leakage from LLM-generated reports. We regenerate reports with label-revealing fields/keywords (e.g., diagnosis/outcome terms) masked, re-train AcuLa using audio + masked reports, and compare downstream performance against AcuLa trained with the original reports across all tasks (T1--T18). Table~\ref{tab:mask_examples} shows representative report snippets before and after masking. As shown in Table~\ref{tab:mask_ablation}, the overlap tasks exhibit a slightly larger but still minor AUROC decrease (mean $\Delta \mathrm{AUROC} = -0.0047$, with a maximum drop of 0.009 on T3), whereas the non-overlap classification tasks show almost no change (mean $\Delta \mathrm{AUROC} = -0.0010$). Regression tasks (T10--T16) remain essentially unchanged, with differences of at most 0.007. Overall, the small and localized degradation on overlap tasks suggests that explicit diagnosis/outcome tokens contribute minimally to alignment, and the largely stable performance across all tasks indicates that AcuLa's gains are not primarily driven by trivial label-text leakage.

\begin{table}[!h]
\centering
\caption{Examples of report text before and after label masking. Diagnosis/outcome keywords are removed to mitigate label leakage, while symptoms and descriptive findings are retained.}
\label{tab:mask_examples}
\footnotesize
\setlength{\tabcolsep}{5pt}
\renewcommand{\arraystretch}{1.2}
\begin{tabular}{p{0.47\linewidth} p{0.47\linewidth}}
\toprule
\textbf{Original (before masking)} & \textbf{Masked (after masking)} \\
\midrule
Cardiac auscultation findings are within normal limits, indicating no abnormalities in heart sounds despite the presence of non-severe pneumonia. &
The record annotation indicates a normal assessment with an event type classified as normal. The findings are within expected parameters with no anomalies noted. \\
\midrule
Auscultation reveals continuous adventitious sounds characterized as wheezes. These findings are consistent with a diagnosis of non-severe pneumonia. &
The patient's lung auscultation recordings demonstrate the presence of both continuous and discontinuous adventitious sounds. \\
\midrule
The patient is a middle-aged female with a positive COVID-19 test presenting with runny or blocked nose, new continuous cough, fatigue, and headache. Respiratory auscultation may reveal mild bronchial breath sounds or scattered crackles related to viral upper respiratory involvement, consistent with her symptom profile. No signs of respiratory distress or wheezing noted. &
This participant is a middle-aged female who does not wear a mask and has never smoked. She reported experiencing a cough, new continuous cough, runny or blocked nose, diarrhea, fatigue, and headache. \\
\midrule
Anterior left chest auscultation reveals no crackles or wheezes in a 3-year-old male diagnosed with bronchiolitis. &
The patient is a 3-year-old male with findings from the anterior left chest. The acquisition mode was sequential/single channel. No crackles or wheezes were noted. \\
\bottomrule
\end{tabular}
\end{table}

\begin{table*}[!h]
\caption{Label-masked report ablation. We mask label-revealing fields/words during report generation (e.g., diagnosis/outcome terms) and re-train AcuLa with audio + masked reports.}
\centering
\footnotesize
\setlength{\tabcolsep}{6pt}
\begin{tabular}{ll|c|c}
\toprule
ID & Task & \textbf{AcuLa (original reports)} & \textbf{AcuLa (masked reports)} \\
\midrule
T1  & Covid (Exhale)         & 0.698  & 0.695 \\
T2  & Covid (Cough)          & 0.730 & 0.724 \\
T3  & Covid (Cough)          & 0.887  & 0.878 \\
T4  & Gender (Cough)         & 0.796  & 0.793 \\
T5  & COPD (Lung)            & 0.826  & 0.819 \\
T6  & Smoker (Cough)         & 0.830  & 0.830 \\
T7  & Gender (Cough)         & 0.845  & 0.845 \\
T8  & Obstructive (Lung)     & 0.752 & 0.750 \\
T9  & COPD severity (Lung)   & 0.710  & 0.708 \\
\midrule
T10 & FVC (Breath)           & 0.865 & 0.865 \\
T11 & FEV1 (Breath)          & 0.742 & 0.744 \\
T12 & FEV1/FVC (Breath)      & 0.127 & 0.127 \\
T13 & FVC (Vowel)            & 0.779 & 0.781 \\
T14 & FEV1 (Vowel)           & 0.725 & 0.724 \\
T15 & FEV1/FVC (Vowel)       & 0.123  & 0.123 \\
T16 & Breathing Rate         & 2.388 & 2.395 \\
\midrule
T17 & Murmur / Healthy        & 0.675 & 0.673 \\
T18 & Symptomatic / Healthy   & 0.647 & 0.644 \\
\bottomrule
\end{tabular}
\label{tab:mask_ablation}
\end{table*}

\clearpage

\section{Baseline Adaptation Results}
\label{sec:baseline_adaptation}
This appendix reports per-task performance under a fair baseline adaptation setting. We fine-tune AudioMAE and CLAP in-domain on the same alignment training audio (train splits only) without using reports, and evaluate all models using the identical frozen linear-probe protocol. Audio-only adaptation improves both encoders compared to their pretrained counterparts, but AcuLa remains consistently stronger on most tasks, particularly for respiratory and cardiac condition inference. These results support that the proposed semantic alignment provides benefits beyond in-domain audio exposure alone. Tables~\ref{tab:fair_resp_tasks}--\ref{tab:fair_cardiac_tasks} provide per-task results for respiratory condition inference (T1--T9, AUROC), lung function estimation (T10--T16, MAE), and cardiac condition inference (T17--T18, AUROC).

\begin{table*}[!h]
\caption{Per-task AUROC ($\uparrow$) on respiratory health condition inference (T1--T9) under fair baseline adaptation. AudioMAE-ft and CLAP-ft are in-domain fine-tuned on the same alignment training audio (train splits only) without reports.}
\centering
\footnotesize
\setlength{\tabcolsep}{6pt}
\begin{tabular}{ll|cc|c}
\toprule
ID & Task & AudioMAE-ft & CLAP-ft & \textbf{AcuLa (Ours)} \\
\midrule
T1 & Covid (Exhale)         & 0.608 & 0.612 & 0.698 \\
T2 & Covid (Cough)          & 0.678 & 0.683 & 0.730 \\
T3 & Covid (Cough)          & 0.671 & 0.679 & 0.887 \\
T4 & Gender (Cough)         & 0.694 & 0.723 & 0.796 \\
T5 & COPD (Lung)            & 0.885 & 0.932 & 0.826 \\
T6 & Smoker (Cough)         & 0.612 & 0.732 & 0.830 \\
T7 & Gender (Cough)         & 0.755 & 0.779 & 0.845 \\
T8 & Obstructive (Lung)     & 0.704 & 0.726 & 0.752 \\
T9 & COPD severity (Lung)   & 0.570 & 0.640 & 0.710 \\
\midrule
\textbf{Avg.} & \textbf{Average} & 0.686 & 0.723 & \textbf{0.786} \\
\bottomrule
\end{tabular}
\label{tab:fair_resp_tasks}
\end{table*}

\begin{table*}[!h]
\caption{Per-task MAE ($\downarrow$) on lung function estimation (T10--T16) under fair baseline adaptation. AudioMAE-ft and CLAP-ft are in-domain fine-tuned on the same alignment training audio (train splits only) without reports.}
\centering
\footnotesize
\setlength{\tabcolsep}{6pt}
\begin{tabular}{ll|cc|c}
\toprule
ID & Task & AudioMAE-ft & CLAP-ft & \textbf{AcuLa (Ours)} \\
\midrule
T10 & FVC (Breath)          & 0.878 & 0.879 & 0.865 \\
T11 & FEV1 (Breath)         & 0.780 & 0.796 & 0.742 \\
T12 & FEV1/FVC (Breath)     & 0.128 & 0.132 & 0.127 \\
T13 & FVC (Vowel)           & 0.845 & 0.856 & 0.779 \\
T14 & FEV1 (Vowel)          & 0.808 & 0.795 & 0.725 \\
T15 & FEV1/FVC (Vowel)      & 0.129 & 0.133 & 0.123 \\
T16 & Breathing Rate        & 2.477 & 2.506 & 2.388 \\
\midrule
\textbf{Avg.} & \textbf{Average} & 0.864 & 0.871 & \textbf{0.821} \\
\bottomrule
\end{tabular}
\label{tab:fair_reg_tasks}
\end{table*}

\begin{table}[!h]
\caption{Per-task AUROC ($\uparrow$) on cardiac condition inference (T17--T18) under fair baseline adaptation. AudioMAE-ft and CLAP-ft are in-domain fine-tuned on the same alignment training audio (train splits only) without reports.}
\centering
\footnotesize
\setlength{\tabcolsep}{6pt}
\begin{tabular}{ll|cc|c}
\toprule
ID & Task & AudioMAE-ft & CLAP-ft & \textbf{AcuLa (Ours)} \\
\midrule
T17 & Murmur / Healthy       & 0.640 & 0.652 & 0.675 \\
T18 & Symptomatic / Healthy  & 0.618 & 0.622 & 0.647 \\
\midrule
\textbf{Avg.} & \textbf{Average} & 0.629 & 0.637 & \textbf{0.661} \\
\bottomrule
\end{tabular}
\label{tab:fair_cardiac_tasks}
\end{table}

\clearpage
\section{Qualitative Retrieval Examples}
\label{apd:retrieve_example}
This section provides representative query cases (Figure \ref{fig:closest_report}). Each row shows the query spectrogram and report with the top three FAISS-retrieved reports. The retrieved texts capture the key findings, demonstrating that the shared embedding space preserves fine-grained diagnostic cues.

\begin{figure*}[!h]
    \captionsetup{position=top,font=small}   
    \caption{Top‑3 clinical reports retrieved for auscultation clips.  
    Left: query spectrogram+reference report.  
    Right: three closest matches returned by our audio–text model.}
    \centering
    \includegraphics[width=0.95\columnwidth]{img/closest.png}
    \label{fig:closest_report}
\end{figure*}

\clearpage

\section{Semantic Alignment Evaluation via Audio--Text Retrieval}
\label{app:retrieval}
We assess cross-modal semantic alignment via bidirectional audio--text retrieval on a held-out set of paired samples. Given a query from one modality (audio or text), the model ranks all candidate items from the other modality by cosine similarity between $\ell_2$-normalized embeddings, and we measure the rank of the true paired item. We report Recall@K (R@K) and median rank (MedR) for both audio$\rightarrow$text and text$\rightarrow$audio retrieval, where candidates are drawn from a pooled test set of size $N \approx 10{,}000$; thus chance performance is extremely low (e.g., random R@10 $\approx 10/N \approx 0.1\%$).

To reduce trivial matching driven by explicit label words, we additionally evaluate retrieval using \emph{label-masked reports}, where fields that directly reveal downstream labels (e.g., diagnosis/outcome keywords such as ``COVID-19 positive'', ``pneumonia'', ``abnormal'') are removed. Masking only alters the report text; the underlying audio--text pairing remains unchanged.

As shown in Table~\ref{tab:retrieval_main}, the aligned embeddings support bidirectional retrieval under this large candidate pool: using original reports, the model attains R@10 of 0.351 (audio$\rightarrow$text) and 0.403 (text$\rightarrow$audio), with MedR of 165 and 128 (i.e., the true pair typically ranks within the top $\sim$1--2\% of candidates). With label-masked reports, performance decreases but remains well above chance (R@10 of 0.311 and 0.356), indicating that retrieval is not solely driven by explicit label words.

\begin{table}[!h]
\caption{Audio$\leftrightarrow$Text retrieval on the held-out test set (candidate pool size $N$). We report Recall@K (higher is better) and median rank (MedR, lower is better). ``Masked reports'' remove label-revealing fields to evaluate non-trivial semantic matching beyond direct label-word alignment.}
\centering
\footnotesize
\setlength{\tabcolsep}{6pt}
\begin{tabular}{lcccc}
\toprule
Setting & R@1 $\uparrow$ & R@5 $\uparrow$ & R@10 $\uparrow$ & MedR $\downarrow$ \\
\midrule
Audio$\rightarrow$Text (original reports) & 0.072 & 0.239 & 0.351 & 165 \\
Audio$\rightarrow$Text (masked reports)   & 0.053 & 0.196 & 0.311 & 229 \\
Text$\rightarrow$Audio (original reports) & 0.087 & 0.274 & 0.403 & 128 \\
Text$\rightarrow$Audio (masked reports)   & 0.066 & 0.236 & 0.356 & 189 \\
\bottomrule
\end{tabular}
\label{tab:retrieval_main}
\end{table}

\section{Performance of a respiratory-only model on respiratory and cardiac audio tasks}
\label{apd:res_only}

We evaluate the performance of a model trained exclusively on respiratory audio data across both respiratory and cardiac tasks. The model maintains strong performance on respiratory-specific tasks (T1-T16), achieving comparable results to our multi-organ trained model. However, performance degrades on cardiac classification tasks (T17-T18), with AUROC scores dropping to approximately 0.6. 

\begin{table*}[!h]
\centering
\caption{Performance on respiratory and cardiac audio tasks using model trained exclusively on respiratory sounds data. T1-T9: respiratory classification (AUROC), T10-T16: lung function estimation (MAE), T17-T18: cardiac classification (AUROC).}
\label{tab:resp_only_combined}
\footnotesize
\setlength{\tabcolsep}{6pt}
\begin{tabular}{ll|c}
\toprule
ID & Task & \textbf{MedGemma-4B} \\
\midrule
T1 & Covid / Non-covid (Exhalation) & 0.695 \\
T2 & Covid / Non-covid (Cough) & 0.748 \\
T3 & Covid / Non-covid (Cough) & 0.885 \\
T4 & Female / Male (Cough) & 0.793 \\
T5 & COPD / Healthy (Lung sounds) & 0.823 \\
T6 & Smoker / Non-smoker (Cough) & 0.838 \\
T7 & Female / Male (Cough) & 0.850 \\
T8 & Obstructive / Healthy (Lung sounds) & 0.745 \\
T9 & COPD severity (Lung sounds) & 0.718 \\
\midrule
T10 & FVC (Breath)       & 0.863 \\
T11 & FEV1 (Breath)      & 0.744 \\
T12 & FEV1/FVC (Breath)  & 0.118 \\
T13 & FVC (Vowel)        & 0.783 \\
T14 & FEV1 (Vowel)       & 0.717 \\
T15 & FEV1/FVC (Vowel)   & 0.124 \\
T16 & Breathing Rate     & 2.391 \\
\midrule
T17 & Murmur / Healthy & 0.605 \\
T18 & Symptomatic / Healthy & 0.597 \\
\bottomrule
\end{tabular}
\end{table*}

\newpage

\section{Respiratory and cardiac audio performance without augmentation}
\label{apd:wo_aug}

We investigate the impact of data augmentation during the alignment phase by evaluating model performance without any augmentation. The results show consistent performance degradation across most tasks compared to our augmented approach. Classification tasks experience AUROC reductions of 0.02-0.05, while regression tasks show increased MAE values, particularly for FEV1/FVC estimation.

\begin{table*}[!h]
\centering
\caption{Performance on respiratory and cardiac audio tasks without data augmentation during alignment. T1-T9: respiratory classification (AUROC), T10-T16: lung function estimation (MAE), T17-T18: cardiac classification (AUROC).}
\label{tab:no_aug_combined}
\footnotesize
\setlength{\tabcolsep}{6pt}
\begin{tabular}{ll|c}
\toprule
ID & Task & \textbf{MedGemma-4B} \\
\midrule
T1 & Covid / Non-covid (Exhalation) & 0.652 \\
T2 & Covid / Non-covid (Cough) & 0.691 \\
T3 & Covid / Non-covid (Cough) & 0.864 \\
T4 & Female / Male (Cough) & 0.782 \\
T5 & COPD / Healthy (Lung sounds) & 0.798 \\
T6 & Smoker / Non-smoker (Cough) & 0.825 \\
T7 & Female / Male (Cough) & 0.833 \\
T8 & Obstructive / Healthy (Lung sounds) & 0.728 \\
T9 & COPD severity (Lung sounds) & 0.673 \\
\midrule
T10 & FVC (Breath)       & 0.908 \\
T11 & FEV1 (Breath)      & 0.779 \\
T12 & FEV1/FVC (Breath)  & 0.135 \\
T13 & FVC (Vowel)        & 0.807 \\
T14 & FEV1 (Vowel)       & 0.745 \\
T15 & FEV1/FVC (Vowel)   & 1.122 \\
T16 & Breathing Rate     & 2.318 \\
\midrule
T17 &  Murmur / Healthy & 0.665 \\
T18 &  Symptomatic / Healthy & 0.627 \\
\bottomrule
\end{tabular}
\end{table*}

\section{Alignment Strategy Analysis}
\label{apd:appendix-alignlayer}

We compare two alignment strategies: aligning only the final transformer layer (Last-L) versus multiple intermediate layers (Multi-L at blocks {3, 6, 9, 12}). Last-L generally outperforms Multi-L, especially in classification tasks. This suggests that aligning only the final layer better preserves hierarchical feature learning and allows more flexible, task-specific representations.

\begin{table*}[ht]
  \centering
\caption{Single–layer vs.\ multi–layer alignment on each downstream task.  
         \textbf{Last-L} = alignment applied \emph{only} to the final transformer block;  
         \textbf{Multi-L} = alignment applied to blocks \{3, 6, 9, last (12)\}, with the four alignment losses averaged. T1-T9: respiratory classification (AUROC), T10-T16: lung function estimation (MAE), T17-T18: cardiac classification (AUROC).}
  \label{tab:align_strategies}
  \footnotesize
  \setlength{\tabcolsep}{6pt}
  \begin{tabular}{ll|cc}
    \toprule
    ID & Task & \textbf{Last-L} & \textbf{Multi-B} \\
    \midrule
    T1 & Covid / Non-covid (Exhalation)           & 0.698 & 0.694 \\
    T2 & Covid / Non-covid (Cough)                & 0.730 & 0.742 \\
    T3 & Covid / Non-covid (Cough)                & 0.887 & 0.879 \\
    T4 & Female / Male (Cough)                    & 0.796 & 0.774 \\
    T5 & COPD / Healthy (Lung sounds)             & 0.826 & 0.820 \\
    T6 & Smoker / Non-smoker (Cough)              & 0.830 & 0.817 \\
    T7 & Female / Male (Cough)                    & 0.845 & 0.818 \\
    T8 & Obstructive / Healthy (Lung sounds)      & 0.752 & 0.770 \\
    T9 & COPD severity (Lung sounds)  & 0.710 & 0.744 \\
    \midrule
    T10 & FVC (Breath)       & 0.865 & 0.840 \\
    T11 & FEV1 (Breath)      & 0.742 & 0.715 \\
    T12 & FEV1/FVC (Breath)  & 0.127 & 0.118 \\
    T13 & FVC (Vowel)        & 0.779 & 0.742 \\
    T14 & FEV1 (Vowel)       & 0.725 & 0.692 \\
    T15 & FEV1/FVC (Vowel)   & 0.123 & 0.125 \\
    T16 & Breathing Rate     & 2.388 & 2.385 \\
    \midrule
    T17 & Murmur / Healthy & 0.675 & 0.660 \\
    T18 & Symptomatic / Healthy & 0.647 & 0.657 \\
    \bottomrule
  \end{tabular}
  \label{tab:multilayer}
\end{table*}
\newpage

\clearpage

\vfill

\end{document}